\documentclass[preprint,showpacs,preprintnumbers,amsmath,amssymb]{revtex4}

\usepackage{graphicx}
\usepackage{dcolumn}
\usepackage{bm}
\usepackage{slashed}
\usepackage{slashbox}

\newcommand{\re}{\mathop{\mathrm{Re}}\nolimits}
\newcommand{\im}{\mathop{\mathrm{Im}}\nolimits}
\newcommand{\notp}{{\slashed{p}}}
\newcommand{\notk}{{\slashed{k}}}

\begin{document}

\preprint{DESY~06-207\hspace{12.55cm} ISSN 0418-9833}
\preprint{MPP-2006-219, NYU-TH/06/10/20\hspace{11.6cm}}

\title{Simple On-Shell Renormalization Framework for the
Cabibbo-Kobayashi-Maskawa Matrix}

\author{Bernd A. Kniehl}
\email{bernd.kniehl@desy.de}
\thanks{permanent address: II. Institut f\"ur Theoretische Physik,
Universit\"at Hamburg, Luruper Chaussee 149, 22761 Hamburg, Germany.}
\author{Alberto Sirlin}
\email{alberto.sirlin@nyu.edu}
\thanks{permanent address: Department of Physics, New York University,
4 Washington Place, New York, New York 10003, USA.}
\affiliation{Max-Planck-Institut f\"ur Physik (Werner-Heisenberg-Institut),
F\"ohringer Ring 6, 80805 Munich, Germany}

\date{\today}

\begin{abstract}
We present an explicit on-shell framework to renormalize the
Cabibbo-Kobayashi-Maskawa (CKM) quark mixing matrix at the one-loop level.
It is based on a novel procedure to separate the external-leg mixing
corrections into gauge-independent self-mass (sm) and gauge-dependent
wave-function renormalization contributions, and to adjust non-diagonal mass
counterterm matrices to cancel all the divergent sm contributions, and also
their finite parts subject to constraints imposed by the hermiticity of the
mass matrices.
It is also shown that the proof of gauge independence and finiteness of the
remaining one-loop corrections to $W\to q_i+\overline{q}_j$ reduces to that in
the unmixed, single-generation case.
Diagonalization of the complete mass matrices leads then to an explicit
expression for the CKM counterterm matrix, which is gauge independent,
preserves unitarity, and leads to renormalized amplitudes that are
non-singular in the limit in which any two fermions become mass degenerate.
\end{abstract}

\pacs{11.10.Gh, 12.15.Ff, 12.15.Lk, 13.38.Be}
\maketitle

\section{\label{sec:one}%
Introduction}

The Cabibbo-Kobayashi-Maskawa (CKM) \cite{cab} quark mixing matrix is one of
the basic pillars of the electroweak sector of the Standard Model (SM).
In fact, the detailed determination of this matrix is one of the major aims of 
recent experiments carried out at the $B$ factories \cite{pdg}, as well as the
objective of a wide range of theoretical studies
\cite{pdg,Czarnecki:2004cw,Marciano:2005ec}.

An important theoretical problem associated with the CKM matrix is its
renormalization.
An early discussion, in the two-generation framework, was presented in
Ref.~\cite{Marciano:1975cn}, which focused mostly on the removal of the
ultraviolet (UV) divergent contributions.
In recent years there have been a number of interesting analyses that address
the renormalization of both the UV-divergent and finite contributions at
various levels of generality and complexity \cite{Denner:1990yz}.

In Ref.~\cite{short}, we outlined an explicit and direct on-shell framework to
renormalize the CKM matrix at the one-loop level, which can be regarded as a
simple generalization of Feynman's approach in Quantum Electrodynamics (QED)
\cite{Feynman:1949zx}.

In the present paper, we present a detailed discussion of this renormalization
framework and of the calculations underpinning its implementation.
We recall that, in QED, the self-energy insertion in an external leg involving
an outgoing fermion is of the form
\begin{eqnarray}
\Delta{\cal M}^{\rm leg}&=&\overline{u}(p)\Sigma(\notp)\frac{1}{\notp-m},
\label{eq:dm}\\
\Sigma(\notp)&=&A+B(\notp-m)+\Sigma_{\rm fin}(\notp),
\label{eq:sig}
\end{eqnarray}
where $u(p)$ is the spinor of the external particle, $\Sigma(\notp)$ the
self-energy, $i(\notp-m)^{-1}$ the particle's propagator, $A$ and $B$
UV-divergent constants, and $\Sigma_{\rm fin}(\notp)$ the finite part that
behaves as $\Sigma_{\rm fin}(\notp)\propto(\notp-m)^2$ in the neighborhood of
$\notp=m$.
The contribution of $A$ to Eq.~(\ref{eq:dm}) exhibits a pole as $\notp\to m$,
while the term proportional to $B$ is regular in this limit and that involving
$\Sigma_{\rm fin}(\notp)$ clearly vanishes.
We may refer to $A$ and $B$ as the ``self-mass'' (sm) and ``wave-function
renormalization'' (wfr) contributions, respectively.
The contribution $A$ is gauge independent and is canceled by the mass
counterterm.
The contribution $B$ is in general gauge dependent but, since the $(\notp-m)$
factor cancels the propagator's singularity, in Feynman's approach it is
combined with the proper vertex diagrams leading to a gauge-independent result.
In other formulations, $B$ in Eq.~(\ref{eq:sig}) is canceled by an explicit
field renormalization counterterm $\delta Z$, which also modifies the
tree-level vertex coupling and, consequently, transfers once more this
contribution to the vertex amplitude.

\begin{figure}[ht]
\begin{center}
\includegraphics[bb=112 626 524 779,width=\textwidth]{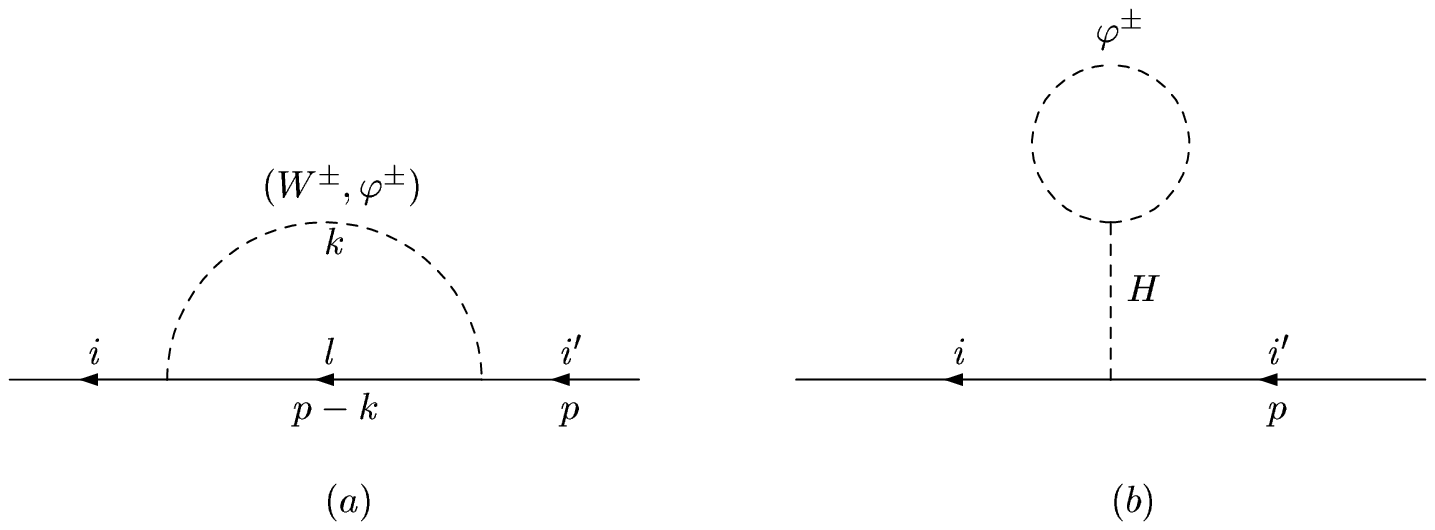}
\end{center}
\caption{\label{fig:one}%
Fermion self-energy diagrams.}
\end{figure}

In the case of the CKM matrix we encounter off-diagonal as well as diagonal
external-leg contributions generated by virtual effects involving $W^\pm$
bosons and charged Goldstone bosons ($\phi^\pm$).
As a consequence, the self-energy corrections to an external leg involving an
outgoing quark is of the form
\begin{equation}
\Delta{\cal M}_{ii^\prime}^{\rm leg}=\overline{u}_i(p)\Sigma_{ii^\prime}(\notp)
\frac{1}{\notp-m_{i^\prime}},
\label{eq:dmii}
\end{equation}
where $i$ denotes the external quark of momentum $p$ and mass $m_i$,
$i^\prime$ the initial virtual quark of mass $m_{i^\prime}$,
$i(\notp-m_{i^\prime})^{-1}$ is the corresponding propagator, and
$\Sigma_{ii^\prime}(\notp)$ the self-energy (see Fig.~\ref{fig:one}).
In Fig.~\ref{fig:one}(b) we have included the tadpole diagram involving a
virtual $\phi^\pm$ boson because its contribution is necessary to remove the
gauge dependence in the diagonal contributions of Fig.~\ref{fig:one}(a).

There are other contributions involving virtual effects of $Z^0$ bosons,
neutral Goldstone bosons ($\phi^0$), photons ($\gamma$), and Higgs bosons
($H$) as well as additional tadpole diagrams, but all of these lead to
diagonal expressions of the usual kind.
An analytic expression for the full result may be found, e.g., in
Ref.~\cite{Kniehl:2000rb}.

In Sec.~\ref{sec:twoa} we analyze in detail the contributions arising from
the diagrams in Fig.~\ref{fig:one}.
After carrying out the Dirac algebra in a way that treats the $i$ and
$i^\prime$ quarks on an equal footing, we find that the 
$\Sigma_{ii^\prime}(\notp)$ contributions can be classified as follows:
(i) terms with a left factor $(\notp-m_i)$;
(ii) terms with a right factor $(\notp-m_{i^\prime})$;
(iii) terms with a left factor $(\notp-m_i)$ and a right factor
$(\notp-m_{i^\prime})$; and
(iv) constant terms not involving $\notp$.

We note that, in Eq.~(\ref{eq:dmii}), $\Sigma_{ii^\prime}(\notp)$ is inserted
between the external-quark spinor $\overline{u}_i(p)$ and the virtual-quark
propagator $i(\notp-m_{i^\prime})^{-1}$.
It follows that off-diagonal contributions of class (i) vanish in
Eq.~(\ref{eq:dmii}), since $(\notp-m_{i^\prime})^{-1}$ is non-singular for
$i^\prime\ne i$, while $\overline{u}_i(p)(\notp-m_i)=0$.
However, there are in general diagonal contributions of class (i), since for
$i^\prime=i$ the factor $(\notp-m_i)$ may cancel against the propagator in
Eq.~(\ref{eq:dmii}).
In contributions of class (ii), the right factor $(\notp-m_{i^\prime})$
cancels the propagator in Eq.~(\ref{eq:dmii}).
In analogy with the cancellation of $\Sigma_{\rm fin}(\notp)$ in
Eqs.~(\ref{eq:dm}) and (\ref{eq:sig}), contributions of class (iii) vanish in
both the diagonal and off-diagonal cases, since the right factor
$(\notp-m_{i^\prime})$ cancels the propagator in Eq.~(\ref{eq:dmii}), and
again $\overline{u}_i(p)(\notp-m_i)=0$.
A common feature of all the non-vanishing contributions to Eq.~(\ref{eq:dmii})
arising from classes (i) and (ii) is that the virtual-quark propagator
$i(\notp-m_{i^\prime})^{-1}$ has been canceled in both the diagonal
$(i^\prime=i)$ and off-diagonal $(i^\prime\ne i)$ cases and, as a consequence,
they are non-singular as $\notp\to m_{i^\prime}$.
Thus, they can be suitably combined with the proper vertex diagrams, in analogy
with $B$ in QED.
In  contrast, the contributions of class (iv) to Eq.~(\ref{eq:dmii}) retain the
virtual-quark propagator $i(\notp-m_{i^\prime})^{-1}$ and are singular in
this limit.

In Sec.~\ref{sec:twoa} we show that, in our formulation, the contributions
to Eq.~(\ref{eq:dmii}) of class (iv) are gauge independent, while those
arising from classes (i) and (ii) contain gauge-dependent pieces.

\begin{figure}[ht]
\begin{center}
\includegraphics[bb=112 634 508 754,width=\textwidth]{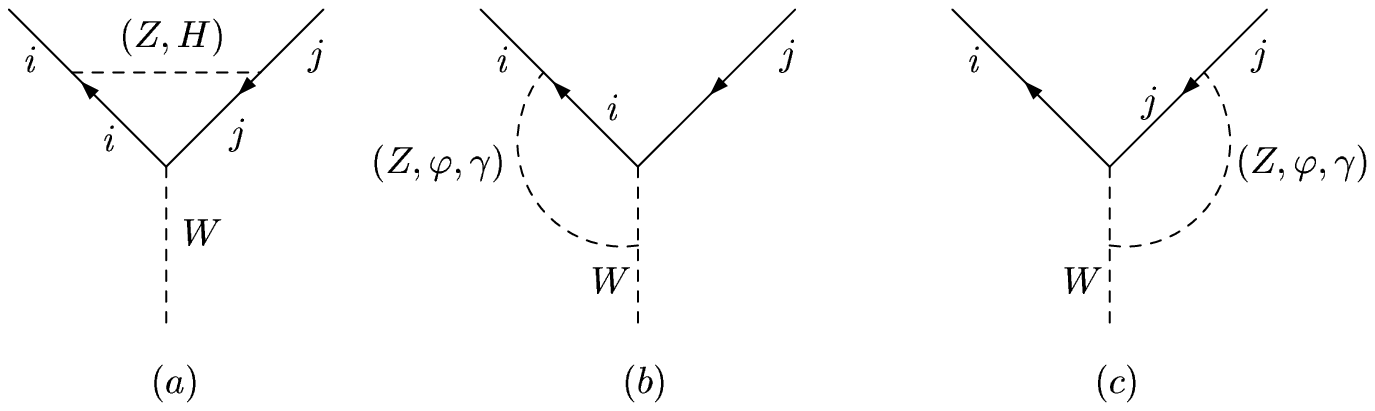}
\end{center}
\caption{\label{fig:two}%
Proper $Wq_i\overline{q}_j$ vertex diagrams.}
\end{figure}

In analogy with the QED case, we identify class (iv) and classes (i) and (ii)
as self-mass (sm) and wave-function renormalization (wfr) contributions,
respectively.
They are listed explicitly in Secs.~\ref{sec:twob} and \ref{sec:twoc}.
In Sec.~\ref{sec:twoc}, we also discuss important simplifications that
occur in the wfr contributions to the physical $W\to q_i+\overline{q}_j$
amplitude.
In particular, we show that the gauge-dependent and the UV-divergent parts of
these contributions depend only on the external-quark masses $m_i$ and $m_j$
and do not involve the CKM matrix elements, except for an overall factor
$V_{ij}$, in analogy with the proper vertex diagrams depicted in 
Fig.~\ref{fig:two}.
This result implies that, once the divergent sm contributions are removed in
the renormalization process, the proof of finiteness and gauge independence of
the remaining one-loop corrections to the $W\to q_i+\overline{q}_j$ amplitude
is the same as in the much simpler case of a hypothetical single generation
made of the $i$ and $j$ quarks with unit CKM coupling.

By contrast, since the sm contributions to Eq.~(\ref{eq:dm}) are proportional
to $(\notp-m_{i^\prime})^{-1}$, they have a structure unsuitable for the
combination with vertex diagrams.
Thus, one expects such terms to be separately gauge independent, as we find.

The plan of this paper is the following.
In Sec.~\ref{sec:two} we evaluate the diagrams depicted in
Fig.~\ref{fig:one} and prove the various properties described above.
In Sec.~\ref{sec:three} we study the cancellation of sm contributions by
suitably adjusting the mass counterterms, subject to restrictions imposed by
hermiticity.
In Sec.~\ref{sec:four} we discuss the diagonalization of the complete mass
matrix, i.e.\ the renormalized plus counterterm mass matrices, and show
explicitly how this procedure generates a CKM counterterm matrix in a manner
that preserves unitarity and gauge independence.
Section~\ref{sec:five} contains our conclusions.

\boldmath
\section{\label{sec:two}%
Evaluation of $\Sigma_{ii^\prime}(\notp)$ and Gauge Independence of the
Self-Mass Contributions}
\unboldmath

In subsection~\ref{sec:twoa} we evaluate the one-loop diagrams of
Fig.~\ref{fig:one}, explain the separation into wfr and sm amplitudes, and
show explicitly the cancellation of gauge dependences in the latter.
Following standard conventions, $\Sigma_{ii^\prime}(\notp)$ is defined as $i$
times the diagrams of Fig.~\ref{fig:one}.

We show how the various contributions can be classified in the categories
(i)--(iv) described in Sec.~\ref{sec:one}.
As explained in Sec.~\ref{sec:one}, terms of class (iii) give a vanishing
contribution to the correction $\Delta{\cal M}_{ii^\prime}^{\rm leg}$
associated with an external leg, while those belonging to classes (i) and (ii)
effectively cancel the virtual-quark propagator $i(\notp-m_{i^\prime})^{-1}$.
They naturally combine with the proper vertex diagrams and are identified with
wfr contributions.
They are generally gauge dependent.
By contrast, in our formulation, the contributions of class (iv) to
$\Delta{\cal M}_{ii^\prime}^{\rm leg}$ are gauge independent and proportional
to $i(\notp-m_{i^\prime})^{-1}$, with a cofactor that is independent of
$\notp$ although it depends on the chiral projectors $a_\pm$.
They are identified with sm contributions.

The sm and wfr contributions to $\Delta{\cal M}_{ii^\prime}^{\rm leg}$ are
given explicitly in subsections~\ref{sec:twob} and \ref{sec:twoc}.

Although the main focus of this paper is the study of the sm contributions,
in Sec.~\ref{sec:twoc} we also digress on the further simplifications of
the wfr contributions that occur in the important $W\to q_i+\overline{q}_j$
amplitude.

\boldmath
\subsection{\label{sec:twoa}%
Evaluation of $\Sigma_{ii^\prime}(\notp)$}
\unboldmath

For definiteness, we first consider the case in which $i$ and $i^\prime$ in
Fig.~\ref{fig:one}(a) are up-type quarks and $l$ is a down-type quark.
Following standard conventions, we denote by $V_{il}$ the CKM matrix element
involving the up-type quark $i$ and the down-type quark $l$.
Simple modifications in other cases are discussed in Sec.~\ref{sec:twod}.

Writing the $W$-boson propagator in the $R_\xi$ gauge as
\begin{equation}
{\cal D}_{\mu\nu}^W=-i\frac{g_{\mu\nu}-k_\mu k_\nu(1-\xi_W)/(k^2-m_W^2\xi_W)}
{k^2-m_W^2},
\label{eq:w}
\end{equation}
where $\xi_W$ is the gauge parameter, we first consider the contribution to
Fig.~\ref{fig:one}(a) of the second, $\xi_W$-dependent term.
We call this contribution $M_{ii^\prime}^{\rm GD}(W)$, where the notation
reminds us that this is the gauge-dependent part of the $W$-boson contribution.
After some elementary algebra, we find
\begin{eqnarray}
M_{ii^\prime}^{\rm GD}(W)&=&\frac{g^2}{2}V_{il}V_{li^\prime}^\dagger(1-\xi_W)
\int_n\frac{1}{\left(k^2-m_W^2\right)\left(k^2-m_W^2\xi_W\right)}
\nonumber\\
&&{}\times a_+\left[-\notk-\notp-m_l+(\notp-m_l)\frac{1}{\notp-\notk-m_l}
(\notp-m_l)\right]a_-,
\end{eqnarray}
where $a_\pm=(1\pm\gamma_5)/2$, $\int_n=\mu^{4-n}\int d^nk/(2\pi)^n$, and
$\mu$ is the 't~Hooft mass scale.
The term proportional to $\notk$ cancels, since the integrand is odd under
$\notk\to-\notk$, and the $m_l$ term cancels because of the chiral
projectors.
We rewrite $\notp a_-$ as follows:
\begin{eqnarray}
2\notp a_-&=&\notp a_-+a_+\notp
\nonumber\\
&=&(\notp-m_i)a_-+a_+(\notp-m_{i^\prime})+m_ia_-+m_{i^\prime}a_+,
\end{eqnarray}
so that the $i$ and $i^\prime$ quarks are treated on an equal footing.
In the terms not involving $m_l$, we employ the unitarity relation,
\begin{equation}
V_{il}V_{li^\prime}^\dagger=\delta_{ii^\prime},
\label{eq:uni}
\end{equation}
and $M_{ii^\prime}^{\rm GD}(W)$ becomes
\begin{eqnarray}
M_{ii^\prime}^{\rm GD}(W)&=&\frac{g^2}{2}(1-\xi_W)
\int_n\frac{1}{\left(k^2-m_W^2\right)\left(k^2-m_W^2\xi_W\right)}
\left\{-\frac{\delta_{ii^\prime}}{2}[a_+(\notp-m_i)+(\notp-m_i)a_-
\right.\nonumber\\
&&{}+\left.m_i]+V_{il}V_{li^\prime}^\dagger
 a_+(\notp-m_l)\frac{1}{\notp-\notk-m_l}(\notp-m_l)a_-\right\}.
\label{eq:mgd}
\end{eqnarray}
The tadpole diagram of Fig.~\ref{fig:one}(b) contributes
\begin{equation}
M_{ii^\prime}^{\rm tad}(\phi)=-\frac{g^2m_i}{4m_W^2}\delta_{ii^\prime}
\int_n\frac{1}{k^2-m_W^2\xi_W}.
\end{equation}
Its combination with the term proportional to $\delta_{ii^\prime}m_i$ in
Eq.~(\ref{eq:mgd}) gives
\begin{equation}
-\frac{g^2m_i}{4m_W^2}\delta_{ii^\prime}\int_n\frac{1}{k^2-m_W^2},
\end{equation}
a gauge-independent amplitude.
Thus,
\begin{eqnarray}
\lefteqn{M_{ii^\prime}^{\rm GD}(W)+M_{ii^\prime}^{\rm tad}(\phi)
=-\frac{g^2m_i}{4m_W^2}\delta_{ii^\prime}\int_n\frac{1}{k^2-m_W^2}}
\nonumber\\
&&{}-\frac{g^2}{4}\delta_{ii^\prime}(1-\xi_W)
\int_n\frac{1}{\left(k^2-m_W^2\right)\left(k^2-m_W^2\xi_W\right)}
[a_+(\notp-m_i)+(\notp-m_i)a_-]
\nonumber\\
&&{}+\frac{g^2}{2}V_{il}V_{li^\prime}^\dagger(1-\xi_W)
\int_n\frac{1}{\left(k^2-m_W^2\right)\left(k^2-m_W^2\xi_W\right)}
a_+(\notp-m_l)\frac{1}{\notp-\notk-m_l}(\notp-m_l)a_-.
\qquad\label{eq:gdtad}
\end{eqnarray}
Using the relations
\begin{eqnarray}
a_+(\notp-m_l)&=&(\notp-m_i)a_-+m_ia_--m_la_+,
\nonumber\\
(\notp-m_l)a_-&=&a_+(\notp-m_{i^\prime})+m_{i^\prime}a_+-m_la_-,
\end{eqnarray}
the last term of Eq.~(\ref{eq:gdtad}) may be written as
\begin{eqnarray}
M_{ii^\prime}^{\rm last}&=&
\frac{g^2}{2}V_{il}V_{li^\prime}^\dagger(1-\xi_W)
\int_n\frac{1}{\left(k^2-m_W^2\right)\left(k^2-m_W^2\xi_W\right)}
[(\notp-m_i)a_-+m_ia_--m_la_+]
\nonumber\\
&&{}\times
\frac{1}{\notp-\notk-m_l}
[a_+(\notp-m_{i^\prime})+m_{i^\prime}a_+-m_la_-].
\label{eq:last}
\end{eqnarray}
On the other hand, the contribution $M_{ii^\prime}(\phi)$ to diagram
\ref{fig:one}(a) arising from the $\phi^\pm$ boson is
\begin{equation}
M_{ii^\prime}(\phi)=
\frac{g^2}{2m_W^2}V_{il}V_{li^\prime}^\dagger
\int_n\frac{1}{k^2-m_W^2\xi_W}
(m_ia_--m_la_+)\frac{1}{\notp-\notk-m_l}
(m_{i^\prime}a_+-m_la_-).
\end{equation}
Its combination with the term proportional to
\begin{equation}
(m_ia_--m_la_+)\frac{1}{\notp-\notk-m_l}(m_{i^\prime}a_+-m_la_-)
\end{equation}
in Eq.~(\ref{eq:last}) leads to a gauge-independent amplitude.

Combining these results, we have
\begin{eqnarray}
\lefteqn{M_{ii^\prime}^{\rm GD}(W)+M_{ii^\prime}^{\rm tad}(\phi)
+M_{ii^\prime}(\phi)
=-\frac{g^2m_i}{4m_W^2}\delta_{ii^\prime}\int_n\frac{1}{k^2-m_W^2}}
\nonumber\\
&&{}+\frac{g^2}{2m_W^2}V_{il}V_{li^\prime}^\dagger
\int_n\frac{1}{k^2-m_W^2}
(m_ia_--m_la_+)\frac{1}{\notp-\notk-m_l}(m_{i^\prime}a_+-m_la_-)
\nonumber\\
&&{}-\frac{g^2}{4}\delta_{ii^\prime}(1-\xi_W)
\int_n\frac{1}{\left(k^2-m_W^2\right)\left(k^2-m_W^2\xi_W\right)}
[a_+(\notp-m_i)+(\notp-m_i)a_-]
\nonumber\\
&&{}+\frac{g^2}{2}V_{il}V_{li^\prime}^\dagger(1-\xi_W)
\int_n\frac{1}{\left(k^2-m_W^2\right)\left(k^2-m_W^2\xi_W\right)}
\left[(\notp-m_i)a_-\frac{1}{\notp-\notk-m_l}a_+(\notp-m_{i^\prime})
\right.
\nonumber\\
&&{}+\left.(\notp-m_i)a_-\frac{1}{\notp-\notk-m_l}(m_{i^\prime}a_+-m_la_-)
+(m_ia_--m_la_+)\frac{1}{\notp-\notk-m_l}a_+(\notp-m_{i^\prime})
\right].
\qquad\label{eq:sum}
\end{eqnarray}
The contribution of the gauge-independent part of the $W$-boson propagator,
i.e.\ the first term in Eq.~(\ref{eq:w}), leads to
\begin{equation}
M_{ii^\prime}^{\rm GI}(W)=
-\frac{g^2}{2}V_{il}V_{li^\prime}^\dagger
\int_n\frac{1}{k^2-m_W^2}
a_+\gamma^\mu\frac{1}{\notp-\notk-m_l}\gamma_\mu a_-.
\label{eq:gi}
\end{equation}
In order to classify the various contributions according to the discussion
of Sec.~\ref{sec:one}, we evaluate the integral that appears in
Eq.~(\ref{eq:gi}) and in the second term of Eq.~(\ref{eq:sum}):
\begin{eqnarray}
K(\notp,m_l)&=&
\int_n\frac{1}{\left(k^2-m_W^2\right)(\notp-\notk-m_l)}
\nonumber\\
&=&-\frac{i}{16\pi^2}\left\{
\notp[\Delta+I(p^2,m_l)-J(p^2,m_l)]+m_l[2\Delta+I(p^2,m_l)]\right\},
\label{eq:k}
\end{eqnarray}
where
\begin{eqnarray}
\Delta&=&\frac{1}{n-4}+\frac{1}{2}[\gamma_E-\ln(4\pi)]+\ln\frac{m_W}{\mu},
\\
\{I(p^2,m_l);J(p^2,m_l)\}
&=&\int_0^1dx\,\{1;x\}\ln\frac{m_l^2x+m_W^2(1-x)-p^2x(1-x)-i\varepsilon}
{m_W^2}.
\label{eq:ij}
\end{eqnarray}
Next, we insert Eq.~(\ref{eq:k}) into the second term of Eq.~(\ref{eq:sum})
and into Eq.~(\ref{eq:gi}) and finally add Eqs.~(\ref{eq:sum}) and
(\ref{eq:gi}).
Treating the terms involving $\notp a_-$ and $\notp a_+$ in the symmetric way
explained before Eq.~(\ref{eq:uni}), evaluating the integral
$\int_n\left(k^2-m_W^2\right)^{-1}$ and employing once more the unitarity
relation~(\ref{eq:uni}) in some of the $m_l$-independent terms, we find that
the complete contribution from Figs.~\ref{fig:one}(a) and (b) can be expressed
in the form:
\begin{eqnarray}
M_{ii^\prime}^{(1)}&=&M_{ii^\prime}^{\rm GD}(W)
+M_{ii^\prime}^{\rm GI}(W)+M_{ii^\prime}^{\rm tad}(\phi)
+M_{ii^\prime}(\phi)
\nonumber\\
&=&\frac{ig^2}{32\pi^2}V_{il}V_{li^\prime}^\dagger
\left\{-m_i\left(1+\frac{m_i^2}{2m_W^2}\Delta\right)
\right.
\nonumber\\
&&{}+\frac{m_l^2}{2m_W^2}
(m_ia_-+m_{i^\prime}a_+)[3\Delta+I(p^2,m_l)+J(p^2,m_l)]
\nonumber\\
&&{}-\left[m_ia_-+m_{i^\prime}a_+
+\frac{m_im_{i^\prime}}{2m_W^2}(m_ia_++m_{i^\prime}a_-)\right]
[I(p^2,m_l)-J(p^2,m_l)]
\nonumber\\
&&{}-\frac{1}{2m_W^2}
\left[m_im_{i^\prime}((\notp-m_i)a_++a_-(\notp-m_{i^\prime}))
+m_l^2((\notp-m_i)a_-+a_+(\notp-m_{i^\prime}))\right]
\nonumber\\
&&{}\times
[\Delta+I(p^2,m_l)-J(p^2,m_l)]
\nonumber\\
&&{}-[(\notp-m_i)a_-+a_+(\notp-m_{i^\prime})]
\left[\Delta+\frac{1}{2}+I(p^2,m_l)-J(p^2,m_l)\right]
\nonumber\\
&&{}+i8\pi^2(1-\xi_W)
\int_n\frac{1}{\left(k^2-m_W^2\right)\left(k^2-m_W^2\xi_W\right)}
[a_+\left(\notp-m_{i^\prime}\right)+(\notp-m_i)a_-]
\nonumber\\
&&{}-i16\pi^2(1-\xi_W)
\int_n\frac{1}{\left(k^2-m_W^2\right)\left(k^2-m_W^2\xi_W\right)}
\nonumber\\
&&{}\times
\left[(\notp-m_i)a_-\frac{1}{\notp-\notk-m_l}a_+(\notp-m_{i^\prime})
+(\notp-m_i)a_-\frac{1}{\notp-\notk-m_l}(m_{i^\prime}a_+-m_la_-)\right.
\nonumber\\
&&{}+\left.\left.
(m_ia_--m_la_+)\frac{1}{\notp-\notk-m_l}a_+(\notp-m_{i^\prime})
\right]\right\}.
\label{eq:one}
\end{eqnarray}
The last two terms in Eq.~(\ref{eq:one}) are gauge dependent and include a
left factor $(\notp-m_i)$  or a right factor $(\notp-m_{i^\prime})$ or both.
Thus, they belong to the classes (i), (ii), or (iii) discussed in
Sec.~\ref{sec:one}.
The integrals in these two terms can readily be evaluated using the identity
\begin{equation}
\frac{1-\xi_W}{\left(k^2-m_W^2\right)\left(k^2-m_W^2\xi_W\right)}
=\frac{1}{m_W^2}\left[\frac{1}{k^2-m_W^2}-\frac{1}{k^2-m_W^2\xi_W}\right]
\label{eq:ide}
\end{equation}
and Eq.~(\ref{eq:k}).
We find
\begin{eqnarray}
&&i8\pi^2(1-\xi_W)\int_n\frac{1}{\left(k^2-m_W^2\right)
\left(k^2-m_W^2\xi_W\right)}
=\Delta-\frac{1}{2}-\xi_W\left(\Delta-\frac{1}{2}+\frac{1}{2}\ln\xi_W\right),
\label{eq:vac}
\\
&&L(\notp,m_l,\xi_W)\equiv i16\pi^2(1-\xi_W)
\int_n\frac{1}{\left(k^2-m_W^2\right)\left(k^2-m_W^2\xi_W\right)
(\notp-\notk-m_l)}
\nonumber\\
&&{}=\frac{1}{m_W^2}\int_0^1dx\,[\notp(1-x)+m_l]\ln
\frac{m_l^2x+m_W^2\xi_W(1-x)-p^2x(1-x)-i\varepsilon}
{m_l^2x+m_W^2(1-x)-p^2x(1-x)-i\varepsilon}.
\label{eq:l}
\end{eqnarray}

If $i$ is an outgoing, on-shell up-type quark, the external-leg amplitude is
obtained by multiplying Eq.~(\ref{eq:one}) on the left by $\overline{u}_i(p)$,
the spinor of the outgoing quark, and on the right by
$i(\notp-m_{i^\prime})^{-1}$, the propagator of the initial virtual quark.
Thus, the relevant amplitude associated with the external leg is
\begin{equation}
\Delta{\cal M}_{ii^\prime}^{\rm leg}=\overline{u}_i(p)M_{ii^\prime}^{(1)}
\frac{i}{\notp-m_{i^\prime}}.
\label{eq:leg}
\end{equation}
This brings about important simplifications.
Using the well-known rules to treat indeterminate factors of the form
$\overline{u}_i(p)(\notp-m_i)(\notp-m_i)^{-1}$ \cite{Feynman:1949zx,schweber},
one readily finds the following identities for both diagonal ($i^\prime=i$)
and off-diagonal ($i^\prime\ne i$) contributions:
\begin{eqnarray}
&&\overline{u}_i(p)[(\notp-m_i)a_\pm+a_\mp(\notp-m_{i^\prime})]
\frac{i}{\notp-m_{i^\prime}}=i\overline{u}_i(p)a_\mp,
\label{eq:idone}\\
&&\overline{u}_i(p)(\notp-m_i)O_1(\notp-m_{i^\prime})
\frac{i}{\notp-m_{i^\prime}}=0,
\label{eq:idtwo}\\
&&\overline{u}_i(p)[(\notp-m_i)a_-L(\notp,m_l,\xi_W)(m_{i^\prime}a_+-m_la_-)
+(m_ia_--m_la_+)L(\notp,m_l,\xi_W)a_+(\notp-m_{i^\prime})]
\nonumber\\
&&\times\frac{i}{\notp-m_{i^\prime}}=i\overline{u}_i(p)(m_ia_--m_la_+)
L(\notp,m_l,\xi_W)a_+,
\label{eq:idthree}
\end{eqnarray}
where $O_1$ is a generic Dirac operator that is regular in the limit
$\notp\to m_{i^\prime}$ and $L(\notp,m_l,\xi_W)$ is the integral defined in
Eq.~(\ref{eq:l}).
These identities tell us that terms in $M_{ii^\prime}^{(1)}$ of class (iii)
give a vanishing contribution to $\Delta{\cal M}_{ii^\prime}^{\rm leg}$
(cf.\ Eq.~(\ref{eq:idtwo})), while those of classes (i) and (ii) combine to
cancel the $(\notp-m_{i^\prime})^{-1}$ factor in Eq.~(\ref{eq:leg})
(cf.\ Eqs.~(\ref{eq:idone}) and (\ref{eq:idthree})).

In the second and third terms of Eq.~(\ref{eq:one}), we expand the functions
$I(p^2,m_l)$ and $J(p^2,m_l)$ about $p^2=m_i^2$.
The lowest-order term, with $p^2$ set equal to $m_i^2$, is independent of
$\notp$ and, therefore, belongs the class (iv).
The same is true of the other contributions in the first two terms of
Eq.~(\ref{eq:one}).
They lead to a multiple of $i(\notp-m_{i^\prime})^{-1}$ in
Eq.~(\ref{eq:leg}) with a cofactor that involves the chiral projectors
$a_\pm$, but is independent of $\notp$.
Thus, they belong to class (iv) and are identified as the sm contributions.
The terms of ${\cal O}\left(p^2-m_i^2\right)$ in the expansions of
$I(p^2,m_l)$ and $J(p^2,m_l)$ give only diagonal contributions ($i^\prime=i$)
to Eq.~(\ref{eq:leg}), belong to class (i)
because $p^2-m_i^2=(\notp-m_i)(\notp+m_i)$, and cancel the
$(\notp-m_i)^{-1}$ factor in Eq.~(\ref{eq:leg}).
Terms of  ${\cal O}\left(\left(p^2-m_i^2\right)^2\right)$ and higher in this
expansion give vanishing contributions to
$\Delta{\cal M}_{ii^\prime}^{\rm leg}$.

As mentioned before, the terms of classes (i) and (ii) in
$M_{ii^\prime}^{(1)}$ (including those generated by the expansions of
$I(p^2,m_l)$ and $J(p^2,m_l)$) are identified as wfr contributions.
In contrast to the sm contributions, they contain gauge-dependent parts
(cf.\ the last two terms in Eq.~(\ref{eq:one})).
Both the sm and wfr contributions contain UV divergences.

\subsection{\label{sec:twob}%
Self-Mass Contributions}

The sm contributions $\Delta{\cal M}_{ii^\prime}^{\rm leg,sm}$ to
the external-leg correction for an outgoing on-shell up-type quark $i$ are
obtained by inserting the first three terms of Eq.~(\ref{eq:one}) with $p^2$
set equal to $m_i^2$ into Eq.~(\ref{eq:leg}):
\begin{eqnarray}
\Delta{\cal M}_{ii^\prime}^{\rm leg,sm}&=&
\frac{g^2}{32\pi^2}V_{il}V_{li^\prime}^\dagger
\overline{u}_i(p)\left\{m_i\left(1+\frac{m_i^2}{2m_W^2}\Delta\right)
\right.
\nonumber\\
&&{}-\frac{m_l^2}{2m_W^2}
(m_ia_-+m_{i^\prime}a_+)
\left[3\Delta+I\left(m_i^2,m_l\right)+J\left(m_i^2,m_l\right)\right]
\nonumber\\
&&{}+\left[m_ia_-+m_{i^\prime}a_+
+\frac{m_im_{i^\prime}}{2m_W^2}(m_ia_++m_{i^\prime}a_-)\right]
\nonumber\\
&&{}\times\left.
\left[I\left(m_i^2,m_l\right)-J\left(m_i^2,m_l\right)\right]\right\}
\frac{1}{\notp-m_{i^\prime}}.
\label{eq:legsm}
\end{eqnarray}
The amplitudes $I\left(m_i^2,m_l\right)$ and $J\left(m_i^2,m_l\right)$,
defined in Eq.~(\ref{eq:ij}), are real except when $m_i=m_t$ corresponding to
an external on-shell top quark.
The diagonal contributions in this case include imaginary parts that cannot be
removed by a mass counterterm, in conjunction with a singular propagator.
The problem arises because, in the usual calculation of its decay rate, the
top quark is treated as an asymptotic state, rather than an unstable particle.
In analogy with the case of the $Z^0$ boson, its proper treatment examines the
resonance region in the virtual propagation of the top quark between its
production and decay vertices.
One finds that, in the narrow-width approximation, in which contributions of
next-to-next-to-leading order are neglected, $\im\Sigma(m_t)$ is related to
the total decay width $\Gamma_t$ by the expression
$\im\Sigma(m_t)=-\Gamma_t[1-\re\Sigma^\prime(m_t)]$ and provides the
$i\Gamma_t$ term in the resonance amplitude.
The latter is proportional to
$i(\notp-m_t+i\Gamma_t)^{-1}[1-\re\Sigma^\prime(m_t)]^{-1}$, where the first
factor is the resonant propagator and the second one the wfr term that
contributes to the top-quark couplings to the external particles in the
production and decay vertices.
Since the imaginary parts of $I\left(m_t^2,m_l\right)$ and
$J\left(m_t^2,m_l\right)$ in the diagonal top-quark contributions are
effectively absorbed in the $i\Gamma_t$ term in the resonance propagator, we
remove them from Eq.~(\ref{eq:legsm}).
Specifically, in the diagonal contributions to Eq.~(\ref{eq:legsm}) involving
an external top quark, $I\left(m_t^2,m_l\right)$ and $J\left(m_t^2,m_l\right)$
are replaced by their real parts.

We see that Eq.~(\ref{eq:legsm}) satisfies the basic properties explained
before:
it is a multiple of the virtual-quark propagator $i(\notp-m_{i^\prime})^{-1}$
with a cofactor that is gauge and momentum independent.
As expected in a chiral theory, it involves the $a_\pm$ projectors.

\subsection{\label{sec:twoc}%
Wave-Function Renormalization Contributions}

For completeness, we exhibit the wfr contributions
$\Delta{\cal M}_{ii^\prime}^{\rm leg,wfr}$ to the external-leg
correction.
They are obtained by inserting the last four terms of Eq.~(\ref{eq:one})
into Eq.~(\ref{eq:leg}), employing the identities of
Eqs.~(\ref{eq:idone})--(\ref{eq:idthree}), and incorporating the diagonal
contributions arising from the expansions of $I(p^2,m_l)$ and $J(p^2,m_l)$ in
the second and third terms of Eq.~(\ref{eq:one}):
\begin{eqnarray}
\Delta{\cal M}_{ii^\prime}^{\rm leg,wfr}&=&
\frac{g^2}{32\pi^2}V_{il}V_{li^\prime}^\dagger\overline{u}_i(p)
\left\{\left[I\left(m_i^2,m_l\right)-J\left(m_i^2,m_l\right)\right]a_+
\right.
\nonumber\\
&&{}+\frac{1}{2m_W^2}(m_im_{i^\prime}a_-+m_l^2a_+)
\left[\Delta+I\left(m_i^2,m_l\right)-J\left(m_i^2,m_l\right)\right]
\nonumber\\
&&{}-\delta_{ii^\prime}\frac{m_i^2m_l^2}{2m_W^2}
\left[I^\prime\left(m_i^2,m_l\right)+J^\prime\left(m_i^2,m_l\right)\right]
+\delta_{ii^\prime}m_i^2\left(1+\frac{m_i^2}{2m_W^2}\right)
\nonumber\\
&&{}\times
\left[I^\prime\left(m_i^2,m_l\right)-J^\prime\left(m_i^2,m_l\right)\right]
+\left[\xi_W\left(\Delta+\frac{1}{2}\ln\xi_W-\frac{1}{2}\right)+1\right]
a_+
\nonumber\\
&&{}-\left.N(m_i,m_l,\xi_W)a_+\right\},
\label{eq:legwfr}
\end{eqnarray}
where
\begin{equation}
\left\{I^\prime\left(m_i^2,m_l\right);J^\prime\left(m_i^2,m_l\right)\right\}
=-\int_0^1\frac{dx\,\{1;x\}x(1-x)}{m_l^2x+m_W^2(1-x)-m_i^2x(1-x)-i\varepsilon}
\end{equation}
are the derivatives of $I(p^2,m_l)$ and $J(p^2,m_l)$ with respect to $p^2$,
evaluated at $p^2=m_i^2$, and
\begin{equation}
N(m_i,m_l,\xi_W)=\frac{1}{m_W^2}\int_0^1dx\,\left[m_i^2(1-x)-m_l^2\right]\ln
\frac{m_l^2x+m_W^2\xi_W(1-x)-m_i^2x(1-x)-i\varepsilon}
{m_l^2x+m_W^2(1-x)-m_i^2x(1-x)-i\varepsilon}.
\end{equation}
The previous to last term in Eq.~(\ref{eq:legwfr}) was obtained by using
Eqs.~(\ref{eq:vac}) and (\ref{eq:idone}), and combining the result with other
$\Delta$-dependent contributions.
The last term in Eq.~(\ref{eq:legwfr}) was obtained by using
Eqs.~(\ref{eq:l}) and (\ref{eq:idthree}), and carrying out some elementary
Dirac algebra.
Employing Eq.~(\ref{eq:uni}) in $m_l$-independent terms, we see that the
UV-divergent part in Eq.~(\ref{eq:legwfr}) is given by
\begin{equation}
\Delta{\cal M}_{ii^\prime}^{\rm leg,wfr,div}
=\frac{g^2}{32\pi^2}V_{il}V_{li^\prime}^\dagger\overline{u}_i(p)
\Delta\left[\frac{m_i^2}{2m_W^2}a_-
+\left(\xi_W+\frac{m_l^2}{2m_W^2}\right)a_+\right],
\label{eq:legdiv}
\end{equation}
which contains both diagonal and off-diagonal pieces.
In particular, the
diagonal part of Eq.~(\ref{eq:legdiv}) contains a gauge-dependent
contribution, while the off-diagonal term is gauge independent.

We now digress on the further simplifications that take place when
Eq.~(\ref{eq:legwfr}) is inserted in the physical $W\to q_i+\overline{q}_j$
amplitude.
In this case, Eq.~(\ref{eq:legwfr}) is multiplied on the right by
$(-ig/\sqrt2)V_{i^\prime j}\gamma^\mu a_-v_j\epsilon_\mu$, where
$v_j$ is the spinor associated with the $\overline{q}_j$ quark and
$\epsilon^\mu$ is the polarization four-vector of the $W$ boson.
Because of the chiral projectors, the contribution of the term proportional to
$(m_im_{i^\prime}/2m_W^2)a_-[\Delta+I-J]$ vanishes.
Next, we note that the first, second, fifth, and sixth terms between curly
brackets in Eq.~(\ref{eq:legwfr}) are independent of $i^\prime$.
Denoting these contributions as $f(m_i,m_l)$ and employing the unitarity
relation~(\ref{eq:uni}), we have
$V_{il}V_{li^\prime}^\dagger V_{i^\prime j}f(m_i,m_l)
=V_{il}\delta_{lj}f(m_i,m_l)=V_{ij}f(m_i,m_j)$.
Thus, the contributions of these terms to the $W\to q_i+\overline{q}_j$
amplitude are proportional to $V_{ij}$ and depend only on the external-fermion
masses $m_i$ and $m_j$.
The same is true of the corresponding contributions arising from the
$\overline{q}_j$ external leg.
We emphasize that this result includes all the gauge-dependent and all the
UV-divergent contributions in Eq.~(\ref{eq:legwfr}).
This important property is shared by the proper vertex diagrams of
Fig.~\ref{fig:two}, which are also proportional to $V_{ij}$ and depend only
on $m_i$ and $m_j$.
As explained in Sec.~\ref{sec:one}, this property implies that, once the
divergent sm contributions are canceled by renormalization, the proof of
finiteness and gauge independence of the remaining one-loop corrections to the
$W\to q_i+\overline{q}_j$ amplitude is the same as in the single-generation 
case.

Although the contributions to the $W\to q_i+\overline{q}_j$ amplitude from the
terms involving $I^\prime\left(m_i^2,m_l\right)$ and
$J^\prime\left(m_i^2,m_l\right)$ in Eq.~(\ref{eq:legwfr}) are not simplified
by the unitarity relations without appealing to suitable approximations, we
note that they are finite and gauge independent.
It is important to point out that the simplifications we encountered in the
$W\to q_i+\overline{q}_j$ amplitude depend crucially on the fact that the wfr
terms cancel the virtual-quark propagator $i(\notp-m_{i^\prime})^{-1}$.

\subsection{\label{sec:twod}%
Other Cases}

Equations~(\ref{eq:legsm}) and (\ref{eq:legwfr}) exhibit the sm and wfr
contributions to the external-leg corrections in the case of an outgoing
on-shell up-type quark $i$.
Here $i^\prime$ labels the initial virtual up-type quark in
Fig.~\ref{fig:one}(a) and $l$ the down-type quark in the loop.

The corresponding expressions for an incoming up-type quark can be gleaned by
multiplying Eq.~(\ref{eq:one}) by $u_{i^\prime}(p)$ on the right and by
$i(\notp-m_i)^{-1}$ on the left.
Interchanging $i$ and $i^\prime$, it is easy to see that the sm contributions
for an incoming up-type quark are obtained from Eq.~(\ref{eq:legsm}) by
substituting $V_{il}V_{li^\prime}^\dagger\to V_{i^\prime l}V_{li}^\dagger$,
interchanging $a_-\leftrightarrow a_+$ between the curly brackets, and
multiplying the resulting expression by $u_i(p)$ on the right and by
$(\notp-m_{i^\prime})^{-1}$ on the left.
Similarly, the wave-function renormalization for an incoming up-type quark is
obtained from Eq.~(\ref{eq:legwfr}) by
substituting $V_{il}V_{li^\prime}^\dagger\to V_{i^\prime l}V_{li}^\dagger$,
interchanging $a_-\leftrightarrow a_+$ between the curly brackets, and
multiplying the resulting expression by $u_i(p)$ on the right.
The expressions for an incoming (outgoing) up-type antiquark are the same as
those for an outgoing (incoming) up-type quark with the substitution
$u_i(p)\to v_i(-p)$, the negative-energy spinor.
In the case of antiquarks, $p$ in these expressions is identified with the 
four-momentum in the direction of the arrows in the Feynman diagrams, which is
minus the four-momentum of the antiparticle.
Finally, the expression for an outgoing down-type quark is obtained from that
of an outgoing up-type quark by substituting
$V_{il}V_{li^\prime}^\dagger\to V_{jl}^\dagger V_{lj^\prime}$, where $j$ and
$j^\prime$ denote the on-shell and virtual down-type quarks, respectively, and
$l$ the up-type quark in the loop.
The other down-type-quark amplitudes are obtained from the corresponding
up-type-quark expressions in a similar manner.

\section{\label{sec:three}%
Mass Renormalization}

In order to generate mass counterterms suitable for the renormalization of
the sm contributions shown in Eq.~(\ref{eq:legsm}), we may proceed as
follows.
In the weak-eigenstate basis, the bare mass matrices $m_0^{\prime Q}$ for the
up- and down-type quarks ($Q=U,D$) are non-diagonal, and the corresponding
terms in the Lagrangian density may be written as
$-\overline{\psi}_R^{\prime Q}m_0^{\prime Q}\psi_L^{\prime Q}+\mbox{h.c.}$,
where $\psi_L^{\prime Q}$ and $\psi_R^{\prime Q}$ are left- and right-handed
column spinors that include the three up-type (or down-type) quarks.
Decomposing  $m_0^{\prime Q}=m^{\prime Q}-\delta m^{\prime Q}$, where
$m^{\prime Q}$ and $\delta m^{\prime Q}$ are identified as the
renormalized and counterterm mass matrices, we envisage a biunitary
transformation of the quark fields that diagonalizes $m^{\prime Q}$,
leading to a renormalized mass matrix $m^Q$ that is diagonal, real and endowed
with positive entries.
The same operation transforms $\delta m^{\prime Q}$ into a new matrix
$\delta m^Q$ which, in general, is non-diagonal.
In the new framework, which we may identify as the mass-eigenstate basis, the
mass term is given by
\begin{equation}
-\overline{\psi}\left(m-\delta m^{(-)}a_--\delta m^{(+)}a_+\right)\psi
=-\overline{\psi}_R\left(m-\delta m^{(-)}\right)\psi_L
-\overline{\psi}_L\left(m-\delta m^{(+)}\right)\psi_R,
\label{eq:mass}
\end{equation}
where $m$ is real, diagonal, and positive, and $\delta m^{(-)}$ and
$\delta m^{(+)}$ are arbitrary non-diagonal matrices subject to the
hermiticity constraint
\begin{equation}
\delta m^{(+)}=\delta m^{(-)\dagger}.
\label{eq:her}
\end{equation}
This constraint follows from the requirement that the mass terms in the
Lagrangian density, displayed in Eq.~(\ref{eq:mass}), must be hermitian.
In order to simplify the notation, we do not exhibit the label $Q$, but it
is understood that Eq.~(\ref{eq:mass}) represents two different mass matrices
involving the up- and down-type quarks.

As is customary, the mass counterterms are included in the interaction
Lagrangian.
Their contribution to Eq.~(\ref{eq:leg}) is given by
\begin{equation}
i\overline{u}_i(p)\left(\delta m_{ii^\prime}^{(-)}a_-
+\delta m_{ii^\prime}^{(+)}a_+\right)\frac{i}{\notp-m_{i^\prime}}.
\label{eq:dmmp}
\end{equation}
We now adjust $\delta m_{ii^\prime}^{(-)}$ and $\delta m_{ii^\prime}^{(+)}$
to cancel, as much as possible, the sm contributions given in
Eq.~(\ref{eq:legsm}).
The cancellation of the UV-divergent parts is achieved by choosing
\begin{eqnarray}
\left(\delta m_{\rm div}^{(-)}\right)_{ii^\prime}&=&
\frac{g^2m_i}{64\pi^2m_W^2}\Delta
\left(\delta_{ii^\prime}m_i^2-3V_{il}V_{li^\prime}^\dagger m_l^2\right),
\nonumber\\
\left(\delta m_{\rm div}^{(+)}\right)_{ii^\prime}&=&
\frac{g^2m_{i^\prime}}{64\pi^2m_W^2}\Delta
\left(\delta_{ii^\prime}m_i^2-3V_{il}V_{li^\prime}^\dagger m_l^2\right),
\end{eqnarray}
It is important to note that
\begin{equation}
\left(\delta m_{\rm div}^{(+)}\right)_{ii^\prime}
=\left(\delta m_{\rm div}^{(-)}\right)_{i^\prime i}^*,
\end{equation}
so that $\delta m_{\rm div}^{(+)}$ and $\delta m_{\rm div}^{(-)}$ satisfy the
hermiticity requirement of Eq.~(\ref{eq:her}).

In order to discuss the cancellation of the finite parts, we call $ii^\prime$
channel the amplitude in which $i$ labels the outgoing, on-shell up-type quark
and ${i^\prime}$ the initial, virtual one (cf.\ Fig.~\ref{fig:one}).
Then the $i^\prime i$ channel is the amplitude in which the roles are
reversed: ${i^\prime}$ is the outgoing, on-shell quark, while $i$ is the
initial, virtual one.

Comparing Eq.~(\ref{eq:legsm}) with Eq.~(\ref{eq:dmmp}), we see that a complete
cancellation of the sm corrections for an outgoing up-type quark or an
incoming up-type antiquark in the $ii^\prime$ channel is achieved by adjusting
the mass counterterms according to
\begin{eqnarray}
\delta m_{ii^\prime}^{(-)}&=&
\frac{g^2m_i}{32\pi^2}\left\{\delta_{ii^\prime}
\left(1+\frac{m_i^2}{2m_W^2}\Delta\right)
-V_{il}V_{li^\prime}^\dagger\frac{m_l^2}{2m_W^2}
\left[3\Delta+I\left(m_i^2,m_l\right)+J\left(m_i^2,m_l\right)\right]\right.
\nonumber\\
&&{}+\left.
V_{il}V_{li^\prime}^\dagger\left(1+\frac{m_{i^\prime}^2}{2m_W^2}\right)
\left[I\left(m_i^2,m_l\right)-J\left(m_i^2,m_l\right)\right]\right\},
\nonumber\\
\delta m_{ii^\prime}^{(+)}&=&
\frac{g^2m_{i^\prime}}{32\pi^2}\left\{\delta_{ii^\prime}
\left(1+\frac{m_i^2}{2m_W^2}\Delta\right)
-V_{il}V_{li^\prime}^\dagger\frac{m_l^2}{2m_W^2}
\left[3\Delta+I\left(m_i^2,m_l\right)+J\left(m_i^2,m_l\right)\right]\right.
\nonumber\\
&&{}+\left.
V_{il}V_{li^\prime}^\dagger\left(1+\frac{m_i^2}{2m_W^2}\right)
\left[I\left(m_i^2,m_l\right)-J\left(m_i^2,m_l\right)\right]\right\}.
\label{eq:mct}
\end{eqnarray}
Once $\delta m_{ii^\prime}^{(-)}$ and $\delta m_{ii^\prime}^{(+)}$ are fixed,
the mass counterterms for the reverse $i^\prime i$ channel are determined by
the hermiticity condition of Eq.~(\ref{eq:her}), to wit
\begin{equation}
\delta m_{i^\prime i}^{(-)}=\delta m_{ii^\prime}^{(+)*},\qquad
\delta m_{i^\prime i}^{(+)}=\delta m_{ii^\prime}^{(-)*}.
\label{eq:mher}
\end{equation}
Since the functions $I$ and $J$ in Eq.~(\ref{eq:legsm}) are evaluated at
$p^2=m_i^2$ in the $ii^\prime$ channel and at $p^2=m_{i^\prime}^2$ in the
$i^\prime i$ channel, we see that the mass counterterms in
Eqs.~(\ref{eq:mct}) and (\ref{eq:mher}) cannot remove completely the sm
contributions in both amplitudes.
Taking into account this restriction, we choose the following renormalization
prescription.

Writing the mass counterterm matrix for the up-type quark in the explicit
form
\begin{equation}
\left(
\begin{array}{ccc}
\delta m_{uu} & \delta m_{uc} & \delta m_{ut} \\
\delta m_{cu} & \delta m_{cc} & \delta m_{ct} \\
\delta m_{tu} & \delta m_{tc} & \delta m_{tt}
\end{array}
\right),
\end{equation}
where 
$\delta m_{ii^\prime}=\delta m_{ii^\prime}^{(-)}a_-
+\delta m_{ii^\prime}^{(+)}a_+$
$(i,i^\prime=u,c,t)$,
we choose $\delta m_{uu}$, $\delta m_{cc}$, and $\delta m_{tt}$ to cancel, as
is customary, all the diagonal contributions in Eq.~(\ref{eq:legsm}).
For the non-diagonal entries, we choose
$\delta m_{uc}$, $\delta m_{ut}$, and $\delta m_{ct}$ to cancel completely the
contributions in Eq.~(\ref{eq:legsm}) corresponding to the $uc$, $ut$, and
$ct$ channels, respectively.
The remaining mass counterterms, $\delta m_{cu}$, $\delta m_{tu}$, and
$\delta m_{tc}$ are then fixed by the hermiticity condition in
Eq.~(\ref{eq:her}).
This implies that the finite parts of the sm corrections in the $cu$,
$tu$, and $tc$ channels are not fully canceled.
However, after the mass renormalization is implemented, the residual
contributions from Eq.~(\ref{eq:legsm}) to the $W\to q_i+\overline{q}_j$
amplitudes are finite, gauge independent, and very small in magnitude (see
Appendix~\ref{sec:A}).
In fact, they are of second (first) order in the small ratios $m_q^2/m_W^2$
($q\ne t)$ when the top quark is not (is) the external particle and,
furthermore, they include small CKM matrix elements.

An analogous approach is followed for the down-type-quark mass counterterms.
We call $j^\prime j$ channel the amplitude involving an incoming, on-shell
down-type quark $j$ and a virtual down-type quark $j^\prime$.
In analogy with Eq.~(\ref{eq:mct}), the complete cancellation of the sm
corrections for an incoming down-type quark (or an outgoing down-type
antiquark) in the $j^\prime j$ channel is implemented by choosing:
\begin{eqnarray}
\delta m_{j^\prime j}^{(-)}&=&
\frac{g^2m_{j^\prime}}{32\pi^2}\left\{\delta_{jj^\prime}
\left(1+\frac{m_j^2}{2m_W^2}\Delta\right)
-V_{j^\prime l}^\dagger V_{lj}\frac{m_l^2}{2m_W^2}
\left[3\Delta+I\left(m_j^2,m_l\right)+J\left(m_j^2,m_l\right)\right]\right.
\nonumber\\
&&{}+\left.
V_{j^\prime l}^\dagger V_{lj}\left(1+\frac{m_j^2}{2m_W^2}\right)
\left[I\left(m_j^2,m_l\right)-J\left(m_j^2,m_l\right)\right]\right\},
\nonumber\\
\delta m_{j^\prime j}^{(+)}&=&
\frac{g^2m_j}{32\pi^2}\left\{\delta_{jj^\prime}
\left(1+\frac{m_j^2}{2m_W^2}\Delta\right)
-V_{j^\prime l}^\dagger V_{lj}\frac{m_l^2}{2m_W^2}
\left[3\Delta+I\left(m_j^2,m_l\right)+J\left(m_j^2,m_l\right)\right]\right.
\nonumber\\
&&{}+\left.
V_{j^\prime l}^\dagger V_{lj}\left(1+\frac{m_{j^\prime}^2}{2m_W^2}\right)
\left[I\left(m_j^2,m_l\right)-J\left(m_j^2,m_l\right)\right]\right\},
\label{eq:mctd}
\end{eqnarray}
where $l$ labels the virtual up-type quark in the self-energy loop.

We emphasize that Eqs.~(\ref{eq:mct}) and (\ref{eq:mctd}) contain all the
off-diagonal sm contributions since they only arise from
Fig.~\ref{fig:one}(a) and the analogous diagrams involving the down-type
quarks.
On the other hand, there are many additional diagonal sm contributions from
other diagrams.

Writing the mass counterterm matrix for the down-type quarks in the form
\begin{equation}
\left(
\begin{array}{ccc}
\delta m_{dd} & \delta m_{ds} & \delta m_{db} \\
\delta m_{sd} & \delta m_{ss} & \delta m_{sb} \\
\delta m_{bd} & \delta m_{bs} & \delta m_{bb}
\end{array}
\right),
\end{equation}
we choose $\delta m_{dd}$, $\delta m_{ss}$, and $\delta m_{bb}$ to cancel the
diagonal sm contributions, and $\delta m_{sd}$, $\delta m_{bd}$, and
$\delta m_{bs}$ to cancel the corresponding off-diagonal terms.
The hermiticity constraint implies then that the finite parts of the sm
contributions are not fully canceled in the reverse $ds$, $db$, and $sb$
channels.
We find that, after the mass renormalization is implemented, the residual
contributions involving the top quark in the self-energy loop are of first
order in the small ratios, while the others are of second order.
Nonetheless, as shown in Appendix~\ref{sec:A}, their contributions to the
$W\to q_i+\overline{q}_j$ amplitudes are also very small.
In particular, the smallness in the $ds$ channel arises because some
contributions are of second order in $m_q^2/m_W^2$  ($q\ne t$) and others are
proportional to $m_s^2/m_t^2$ with very small CKM coefficients.

We note that, in these renormalization prescriptions, the residual sm
contributions are convergent in the limit $m_{i^\prime}\to m_i$ or
$m_{j^\prime}\to m_j$, since the singularities of the virtual propagators
$i(\notp-m_{i^\prime})^{-1}$ and $i(\notp-m_{j^\prime})^{-1}$ are canceled, a
characteristic property of wfr contributions.
Thus, these residual sm terms can be regarded as additional finite and
gauge-independent contributions to wave-function renormalization that happen
to be numerically very small.

It is also interesting to note that these renormalization prescriptions imply
that the sm contributions are fully canceled when the $u$ or $d$ quarks or
antiquarks are the external, on-shell particles.
This is of special interest since $V_{ud}$, the relevant parameter in the
$W\to u+\overline{d}$ amplitude, is by far the most accurately measured CKM
matrix element \cite{Czarnecki:2004cw,Marciano:2005ec}.

\section{\label{sec:four}%
Diagonalization of the Mass Counterterms and Derivation of the CKM Counterterm
Matrix}

In Sec.~\ref{sec:three}, we showed explicitly how the UV-divergent parts of
the one-loop sm contributions associated with external quark legs [cf.\
Fig.~\ref{fig:one}(a)] can be canceled by suitably adjusting the non-diagonal
mass counterterm matrix.
By imposing on-shell renormalization conditions, we also showed how the finite
parts of such contributions can be canceled up to the constraints imposed by
the hermiticity of the mass matrix.
We also recall that, in our formulation, the sm contributions and,
consequently, also the mass counterterms are explicitly gauge independent.

In this section, we discuss the diagonalization of the complete mass matrix of
Eq.~(\ref{eq:mass}), which includes the renormalized and counterterm mass
matrices.
We show how this procedure generates a CKM counterterm matrix that
automatically satisfies the basic properties of gauge independence and
unitarity.

Starting with Eq.~(\ref{eq:mass}), we implement a biunitary transformation
that diagonalizes the matrix $m-\delta m^{(-)}$.
Specifically, we consider the transformations
\begin{eqnarray}
\psi_L&=&U_L\hat\psi_L,
\label{eq:pl}\\
\psi_R&=&U_R\hat\psi_R,
\label{eq:pr}
\end{eqnarray}
and choose the unitary matrices $U_L$ and $U_R$ so that
\begin{equation}
U_R^\dagger\left(m-\delta m^{(-)}\right)U_L={\cal D},
\label{eq:d}
\end{equation}
where $\cal D$ is diagonal and real.
From Eq.~(\ref{eq:d}), it follows that
\begin{equation}
U_L^\dagger\left(m-\delta m^{(-)\dagger}\right)\left(m-\delta m^{(-)}\right)
U_L={\cal D}^2,
\end{equation}
which, through ${\cal O}(g^2)$, reduces to
\begin{equation}
U_L^\dagger\left(m^2-m\delta m^{(-)}-\delta m^{(-)\dagger}m\right)U_L
={\cal D}^2.
\end{equation}
Writing $U_L=1+ih_L$, where $h_L$ is hermitian and of ${\cal O}(g^2)$, we have
\begin{equation}
m^2+i(m^2h_L-h_Lm^2)-m\delta m^{(-)}-\delta m^{(-)\dagger}m={\cal D}^2,
\label{eq:hl}
\end{equation}
where we have neglected terms of ${\cal O}(g^4)$.
Recalling that, in our formulation, $m$ is diagonal (cf.\
Sec.~\ref{sec:three}) and taking the $ii^\prime$ component,
Eq.~(\ref{eq:hl}) becomes
\begin{equation}
m_i^2\delta_{ii^\prime}+i\left(m_i^2-m_{i^\prime}^2\right)(h_L)_{ii^\prime}
-m_i\delta m_{ii^\prime}^{(-)}-\delta m_{ii^\prime}^{(-)\dagger}m_{i^\prime}
={\cal D}_i^2\delta_{ii^\prime}.
\label{eq:dis}
\end{equation}
For diagonal terms, with $i=i^\prime$, the term proportional to
$(h_L)_{ii^\prime}$ does not contribute.
Furthermore, Eq.~(\ref{eq:mct}) tells us that
$\delta m_{ii}^{(-)}=\delta m_{ii}^{(+)}$.
Consequently, for diagonal elements of the mass counterterm matrix, one has
$\delta m_{ii}^{(-)}a_-+\delta m_{ii}^{(+)}a_+=\delta m_i$, where
$\delta m_i=\delta m_{ii}^{(-)}=\delta m_{ii}^{(+)}$.
We note that the hermiticity condition of Eq.~(\ref{eq:mher}) implies that
$\delta m_i$ is real.
Therefore, for $i=i^\prime$, Eq.~(\ref{eq:dis}) reduces to
$m_i^2-2m_i\delta m_i={\cal D}_i^2$ or, equivalently, through ${\cal O}(g^2)$,
to
\begin{equation}
{\cal D}_i=m_i-\delta m_i.
\label{eq:di}
\end{equation}
In order to satisfy Eq.~(\ref{eq:dis}) for $i\ne i^\prime$, we need to cancel
the off-diagonal contributions
$m_i\delta m_{ii^\prime}^{(-)}+\delta m_{ii^\prime}^{(-)\dagger}m_{i^\prime}$.
This is achieved by adjusting the non-diagonal elements of $h_L$ according to
\begin{equation}
i(h_L)_{ii^\prime}=\frac{m_i\delta m_{ii^\prime}^{(-)}
+\delta m_{ii^\prime}^{(+)}m_{i^\prime}}{m_i^2-m_{i^\prime}^2}
\qquad (i\ne i^\prime),
\label{eq:hlii}
\end{equation}
where we have employed the hermiticity relation of Eq.~(\ref{eq:her}).
Since the diagonal elements $(h_L)_{ii}$ do not contribute to
Eq.~(\ref{eq:dis}), it is convenient to choose $(h_L)_{ii}=0$.
In Appendix~\ref{sec:B}, we show that the alternative selection
$(h_L)_{ii}\ne0$ has no physical effect on the $Wq_i\overline{q}_j$
interactions.

Returning to Eq.~(\ref{eq:d}) and writing $U_R=1+ih_R$, one finds that $h_R$
is obtained from $h_L$ by substituting
$\delta m^{(-)}\leftrightarrow\delta m^{(+)}$ in Eq.~(\ref{eq:hlii}).
Thus,
\begin{equation}
i(h_R)_{ii^\prime}=\frac{m_i\delta m_{ii^\prime}^{(+)}
+\delta m_{ii^\prime}^{(-)}m_{i^\prime}}{m_i^2-m_{i^\prime}^2}
\qquad (i\ne i^\prime).
\label{eq:hrii}
\end{equation}
In fact, substituting $U_L=1+ih_L$ and $U_R=1+ih_R$ in Eq.~(\ref{eq:d}) and
employing Eqs.~(\ref{eq:hlii}) and (\ref{eq:hrii}), one readily verifies that
the l.h.s.\ of Eq.~(\ref{eq:d}) is indeed diagonal through ${\cal O}(g^2)$.
Furthermore, one recovers Eq.~(\ref{eq:di}).

The above analysis is carried out separately to diagonalize the mass matrices
of the up- and down-type quarks.
Thus, we obtain two pairs of $h_L$ and $h_R$ matrices: $h_L^U$ and $h_R^U$ for
the up-type quarks and $h_L^D$ and $h_R^D$ for the down-type quarks.

Next, we analyze the effect of transformation (\ref{eq:pl}) on the
$Wq_i\overline{q}_j$ interaction.
Following standard conventions, the latter is given by
\begin{equation}
{\cal L}_{Wq_i\overline{q}_j}=-\frac{g_0}{\sqrt2}\overline{\psi}_i^U
V_{ij}\gamma^\lambda a_-\psi_j^DW_\lambda+\mbox{h.c.},
\end{equation}
where $\psi_i^U$ ($i=u,c,t$) and $\psi_j^D$ ($j=d,s,b$) are the fields
of the up- and down-type quarks, respectively, $W_\lambda$ is the field that
annihilates a $W^+$ boson or creates a $W^-$ boson, $g_0$ is the bare
SU(2)$_L$ coupling, and $V_{ij}$ are the elements of the unitary CKM matrix.
Alternatively, in matrix notation, we have
\begin{equation}
{\cal L}_{Wq_i\overline{q}_j}=-\frac{g_0}{\sqrt2}\overline{\psi}_L^U
V\gamma^\lambda\psi_L^DW_\lambda+\mbox{h.c.}.
\label{eq:lwqq}
\end{equation}
It is important to note that, in the formulation of this paper, in which the
UV-divergent sm terms are canceled by the mass counterterms and the proof of
finiteness of the other contributions to the $W\to q_i+\overline{q}_j$
amplitude after the renormalization of $g_0$ is the same as in the unmixed
case (cf.\ Sec.~\ref{sec:twoc}), $V_{ij}$ are finite quantities.

Inserting Eq.~(\ref{eq:pl}) in Eq.~(\ref{eq:lwqq}), we find, through terms of
${\cal O}(g^2)$, that
\begin{equation}
{\cal L}_{Wq_i\overline{q}_j}=-\frac{g_0}{\sqrt2}\overline{\hat\psi}_L^U
(V-\delta V)\gamma^\lambda\hat\psi_L^DW_\lambda+\mbox{h.c.},
\label{eq:hc}
\end{equation}
where
\begin{equation}
\delta V=i\left(h_L^UV-Vh_L^D\right).
\label{eq:dv}
\end{equation}
One readily verifies that $V-\delta V$ satisfies the unitarity condition
through terms of ${\cal O}(g^2)$, namely
\begin{equation}
(V-\delta V)^\dagger(V-\delta V)=1+{\cal O}(g^4).
\end{equation}
Since $V$ is finite and unitary, it is identified with the renormalized CKM
matrix.
On the other hand, in the $(\hat\psi_L,\hat\psi_R)$ basis, in which the
complete quark mass matrices are diagonal, $\delta V$ and $V_0=V-\delta V$
represent the counterterm and bare CKM matrices, respectively.

We now show explicitly that the $ih_L^UV$ term in $\delta V$ leads to the same
off-diagonal contribution to the $W\to q_i+\overline{q}_j$ amplitude as the
insertion of the mass counterterms $\delta m^{U(-)}$ and $\delta m^{U(+)}$ in
the external up-type-quark line.
Indeed, the $ih_L^UV$ contribution is given by
\begin{equation}
{\cal M}(ih_L^UV)=\frac{ig}{\sqrt2}\overline{u}_ii
\left(h_L^U\right)_{ii^\prime}V_{i^\prime j}\gamma^\lambda a_-v_j
\epsilon_\lambda,
\label{eq:mihv}
\end{equation}
where, again, $u_i$ and $v_j$ are the external up- and down-type-quark spinors,
respectively, and $\epsilon_\lambda$ is the $W$-boson polarization four-vector.
Inserting Eq.~(\ref{eq:hlii}), Eq.~(\ref{eq:mihv}) becomes
\begin{equation}
{\cal M}(ih_L^UV)=\frac{ig}{\sqrt2}\overline{u}_i
\frac{m_i^U\delta m_{ii^\prime}^{U(-)}
+\delta m_{ii^\prime}^{U(+)}m_{i^\prime}^U}{\left(m_i^U\right)^2
-\left(m_{i^\prime}^U\right)^2}
V_{i^\prime j}\gamma^\lambda a_-v_j\epsilon_\lambda,
\label{eq:mihv1}
\end{equation}
where it is understood that $i\ne i^\prime$ and the label $Q=U,D$, which we
had suppressed from Eq.~(\ref{eq:mass}) through Eq.~(\ref{eq:hrii}), is again
displayed.
On the other hand, the off-diagonal mass counterterm insertion in the external
up-type-quark line is given by
\begin{equation}
{\cal M}\left(\delta m^{U(-)},\delta m^{U(+)}\right)
=-\frac{ig}{\sqrt2}\overline{u}_ii\left(\delta m_{ii^\prime}^{U(-)}a_-
+\delta m_{ii^\prime}^{U(+)}a_+\right)\frac{i}{\notp-m_{i^\prime}^U}
V_{i^\prime j}\gamma^\lambda a_-v_j\epsilon_\lambda.
\label{eq:mmm}
\end{equation}
Rationalizing the propagator $i(\notp-m_{i^\prime}^U)^{-1}$, one finds after
some elementary algebra that Eq.~(\ref{eq:mihv1}) coincides with
Eq.~(\ref{eq:mmm}).
An analogous calculation shows that the $-iVh_L^D$ term in $\delta V$ leads to
the same off-diagonal contribution to the  $W\to q_i+\overline{q}_j$ amplitude
as the insertion of the mass counterterms $\delta m^{D(-)}$ and
$\delta m^{D(+)}$ in the external down-type-quark line.
Since the mass counterterms are adjusted to cancel the off-diagonal sm
contributions to the extent allowed by the hermiticity of the mass matrix, the
same is true of the CKM counterterm matrix $\delta V$.
In particular, $\delta V$ fully cancels the UV-divergent part of the
off-diagonal sm contributions.
As mentioned above, in the formulation of this section, the complete mass
matrix is diagonal, with elements of the form given in Eq.~(\ref{eq:di}),
where $m_i$ are the renormalized masses and $\delta m_i$ the corresponding
mass counterterms.
The quantities $\delta m_i$ are then adjusted to fully cancel the diagonal sm
corrections in the external legs, in analogy with QED.
As also explained above, the additional UV divergences arising from the wfr
contributions, proper vertex diagrams, and renormalization of $g_0$ cancel
among themselves as in the single-generation case.

For completeness, we explicitly exhibit the counterterm of the CKM matrix in
component form:
\begin{eqnarray}
\delta V_{ij}&=&i\left[\left(h_L^U\right)_{ii^\prime}V_{i^\prime j}
-V_{ij^\prime}\left(h_L^D\right)_{j^\prime j}\right]
\nonumber\\
&=&\frac{m_i^U\delta m_{ii^\prime}^{U(-)}
+\delta m_{ii^\prime}^{U(+)}m_{i^\prime}^U}{\left(m_i^U\right)^2
-\left(m_{i^\prime}^U\right)^2}V_{i^\prime j}
-V_{ij^\prime}\frac{m_{j^\prime}^D\delta m_{j^\prime j}^{D(-)}
+\delta m_{j^\prime j}^{D(+)}m_j^D}{\left(m_{j^\prime}^D\right)^2
-\left(m_j^D\right)^2},
\label{eq:dvii}
\end{eqnarray}
where we have used Eqs.~(\ref{eq:hlii}), (\ref{eq:hrii}), and (\ref{eq:dv})
and it is understood that $i\ne i^\prime$ in the first term and
$j^\prime\ne j$ in the second one.

We note that Eq.~(\ref{eq:dvii}) involves contributions proportional to
$\left(m_i^U-m_{i^\prime}^U\right)^{-1}$ and
$\left(m_{j^\prime}^D-m_j^D\right)^{-1}$, which would become very large if the
masses of different flavors were nearly degenerate.
This is to be expected, since the role of these counterterms is precisely to
cancel the analogous sm contributions to Eq.~(\ref{eq:dmii}) arising from
Fig.~\ref{fig:one}, so that the renormalized expressions are indeed free from
such singular behavior.

It is important to emphasize that, in this formulation, both the renormalized
CKM matrix $V$ and its bare counterpart $V_0=V-\delta V$ are explicitly gauge
independent and satisfy the unitarity constraints $V^\dagger V=1$ and
$V_0^\dagger V_0=1$, respectively, through the order of the calculation.
The explicit construction of the CKM counterterm matrix, as given in 
Eqs.~(\ref{eq:dv}) and (\ref{eq:dvii}), satisfying this important property, is
the main result of this section.

\section{\label{sec:five}%
Conclusions}

In this paper we have presented a natural on-shell framework to renormalize the
CKM matrix at the one-loop level.
We have shown the gauge independence of the sm contributions and discussed
their cancellation in two equivalent formulations:
the first one involves non-diagonal mass counterterms, while the second one is
based on a CKM counterterm matrix.
We have also established the important fact that the proof of gauge
independence and finiteness of the remaining one-loop corrections to the
$W\to q_i+\overline{q}_j$ amplitude can be reduced to the single-generation
case.
The analysis has led us to an explicit expression for the CKM counterterm
matrix $\delta V_{ij}$, given in Eq.~(\ref{eq:dvii}), that satisfies the basic
property of gauge independence and is consistent with the unitarity of both
$V_0=V-\delta V$ and $V$, the bare and renormalized CKM matrices.
Furthermore, it leads to renormalized amplitudes that are non-singular in the
limit in which any two fermions become mass degenerate.
Because $V$ is finite, gauge independent, and unitary, its elements can be
identified with the experimentally measured CKM matrix elements.

\begin{acknowledgments}

We are grateful to the Max Planck Institute for Physics in Munich for the
hospitality during a visit when this manuscript was finalized.
The work of B.A.K. was supported in part by the German Research Foundation
through the Collaborative Research Center No.\ 676 {\it Particles, Strings and
the Early Universe---the Structure of Matter and Space-Time}.
The work of A.S. was supported in part by the Alexander von Humboldt
Foundation through the Humboldt Reseach Award No.\ IV~USA~1051120~USS and by
the National Science Foundation through Grant No.\ PHY-0245068.

\end{acknowledgments}

\begin{appendix}

\boldmath
\section{\label{sec:A}%
Residual Self-Mass Corrections $C_{ij}$}
\unboldmath

In this appendix we evaluate the finite and gauge-independent residual
contributions $-C_{ij}\overline{u}_i\gamma^\lambda a_-v_j\epsilon_\lambda$ to
the $W\to q_i + \overline{q}_j$ amplitude that are not removed in our mass
renormalization prescription due to the restrictions imposed by the
hermiticity of the mass matrices.
Inserting Eq.~(\ref{eq:legsm}) and its counterpart for down-quark matrices in
the expression for the $W\to q_i+\overline{q}_j$ amplitude and implementing
our mass renormalization subtractions, we find the residual sm corrections
$C_{ij}$ to be
\begin{eqnarray}
C_{ij}&=&\frac{g^2}{32\pi^2}\left\{
\frac{V_{il}V_{li^\prime}^\dagger V_{i^\prime j}}{m_i^2-m_{i^\prime}^2}
\left[\left(m_i^2+m_{i^\prime}^2+\frac{m_i^2m_{i^\prime}^2}{m_W^2}\right)
(I(p^2,m_l)-J(p^2,m_l))\right.\right.
\nonumber\\
&&{}-\left.\frac{m_l^2}{2m_W^2}\left(m_i^2+m_{i^\prime}^2\right)
(I(p^2,m_l)+J(p^2,m_l))\right]_{p^2=m_{i^\prime}^2}^{p^2=m_i^2}
\nonumber\\
&&{}+\frac{V_{ij^\prime}V_{j^\prime k}^\dagger V_{kj}}{m_j^2-m_{j^\prime}^2}
\left[\left(m_j^2+m_{j^\prime}^2+\frac{m_j^2m_{j^\prime}^2}{m_W^2}\right)
(I(p^2,m_k)-J(p^2,m_k))\right.
\nonumber\\
&&{}-\left.\left.\frac{m_k^2}{2m_W^2}\left(m_j^2+m_{j^\prime}^2\right)
(I(p^2,m_k)+J(p^2,m_k))\right]_{p^2=m_{j^\prime}^2}^{p^2=m_j^2}\right\},
\label{eq:shift}
\end{eqnarray}
where the $l$ and $k$ summations are over $l=d,s,b$ and $k=u,c,t$, and it is
understood that only terms with $(i,i^\prime)=(c,u),(t,u),(t,c)$ or
$(j^\prime,j)=(d,s),(d,b),(s,b)$ are included.

For the reader's convenience, we list compact analytic results for the
functions $I(p^2,m_l)$ and $J(p^2,m_l)$ defined in Eq.~(\ref{eq:ij}):
\begin{eqnarray}
I(p^2,m_l)&=&-2+\frac{p^2+m_l^2-m_W^2}{2p^2}\ln\frac{m_l^2}{m_W^2}
-2\frac{m_lm_W}{p^2}f\left(\frac{p^2-m_l^2-m_W^2}{2m_lm_W}\right),
\nonumber\\
J(p^2,m_l)&=&\frac{1}{2p^2}\left[-m_l^2+m_W^2+m_l^2\ln\frac{m_l^2}{m_W^2}
+\left(p^2-m_l^2+m_W^2\right)I(p^2,m_l)\right],
\label{eq:ana}
\end{eqnarray}
where
\begin{equation}
f(x)=
\begin{cases}
\sqrt{x^2-1}\cosh^{-1}(-x) & \text{if $x\le-1$,} \\
-\sqrt{1-x^2}\cos^{-1}(-x) & \text{if $-1<x\le1$,} \\
\sqrt{x^2-1}\left(-\cosh^{-1}x+i\pi\right) & \text{if $x>1$.}
\end{cases}
\end{equation}
In practical applications of Eq.~(\ref{eq:ana}), one encounters strong
numerical cancellations between the various terms when $|p^2|\ll m_W^2$.
It is then advantageous to employ the expansions of $I(p^2,m_l)$ and
$J(p^2,m_l)$ in $p^2$ about $p^2=0$,
\begin{eqnarray}
I(p^2,m_l)&=&-1+\frac{m_l^2}{m_l^2-m_W^2}\ln\frac{m_l^2}{m_W^2}
+\frac{p^2}{\left(m_l^2-m_W^2\right)^2}
\left(-\frac{m_l^2+m_W^2}{2}+\frac{m_l^2m_W^2}{m_l^2-m_W^2}
\ln\frac{m_l^2}{m_W^2}\right)
\nonumber\\
&&{}+{\cal O}\left((p^2)^2\right),
\nonumber\\
J(p^2,m_l)&=&\frac{1}{2\left(m_l^2-m_W^2\right)}
\left(\frac{-m_l^2+3m_W^2}{2}
+\frac{m_l^2\left(m_l^2-2m_W^2\right)}{m_l^2-m_W^2}
\ln\frac{m_l^2}{m_W^2}\right)
\nonumber\\
&&{}+\frac{p^2}{\left(m_l^2-m_W^2\right)^3}
\left(\frac{-m_l^4+5m_l^2m_W^2+2m_W^4}{6}
-\frac{m_l^2m_W^4}{m_l^2-m_W^2}\ln\frac{m_l^2}{m_W^2}\right)
\nonumber\\
&&+{\cal O}\left((p^2)^2\right).
\end{eqnarray}

The standard parameterization of the CKM matrix, in terms of the three angles
$\theta_{12}$, $\theta_{23}$, and $\theta_{13}$ and the phase $\delta$, reads
\cite{pdg}:
\begin{equation}
V=\left(
\begin{array}{ccc}
V_{ud} & V_{us} & V_{ub} \\
V_{cd} & V_{cs} & V_{cb} \\
V_{td} & V_{ts} & V_{tb}
\end{array}
\right)
=\left(
\begin{array}{ccc}
c_{12}c_{13} & s_{12}c_{13} & s_{13}e^{-i\delta} \\
-s_{12}c_{23}-c_{12}s_{23}s_{13}e^{i\delta} &
c_{12}c_{23}-s_{12}s_{23}s_{13}e^{i\delta} & s_{23}c_{13} \\
s_{12}s_{23}-c_{12}c_{23}s_{13}e^{i\delta} &
-c_{12}s_{23}-s_{12}c_{23}s_{13}e^{i\delta} & c_{23}c_{13}
\end{array}
\right),
\label{eq:vckm}
\end{equation}
where $s_{ij}=\sin\theta_{ij}$ and $c_{ij}=\cos\theta_{ij}$.
An equivalent set of four real parameters are $\lambda$, $A$,
$\overline{\rho}$, and $\overline{\eta}$, which are related to
$\theta_{12}$, $\theta_{23}$, $\theta_{13}$, and $\delta$ as \cite{pdg}
\begin{eqnarray}
s_{12}&=&\lambda,\nonumber\\
s_{23}&=&A\lambda^2,\nonumber\\
s_{13}e^{i\delta}&=&\frac{A\lambda^3(\overline{\rho}+i\overline{\eta})
\sqrt{1-A^2\lambda^4}}{\sqrt{1-\lambda^2}
\left[1-A^2\lambda^4(\overline{\rho}+i\overline{\eta})\right]}.
\label{eq:wolf}
\end{eqnarray}

In our numerical evaluation of Eq.~(\ref{eq:shift}), we identify
$g^2/(4\pi)=\hat\alpha(m_Z)/\sin^2\hat\theta_W(m_Z)$ and employ the values
$\hat\alpha(m_Z)=1/127.918$ and $\sin^2\hat\theta_W(m_Z)=0.23122$ \cite{pdg}.
We take the $W$-boson mass to be $m_W=80.403$~GeV \cite{pdg}. 
As for the quark masses, we use the values $m_u=62$~MeV, $m_d=83$~MeV,
$m_s=215$~MeV, $m_c=1.35$~GeV, $m_b=4.5$~GeV \cite{Czarnecki:2004cw} and
$m_t=172.7$~GeV \cite{pdg}; in the case of the lighter quarks, these
correspond to effective masses that are especially appropriate for electroweak
analyses like ours.
We evaluate the CKM matrix elements from Eqs.~(\ref{eq:vckm}) and
(\ref{eq:wolf}) using the values $\lambda=0.2272$, $A=0.818$,
$\overline{\rho}=0.221$, and $\overline{\eta}=0.340$ \cite{pdg}.

\begin{table}
\begin{center}
\caption{Residual self-mass corrections $C_{ij}$ as evaluated from
Eq.~(\ref{eq:shift}) in the form $(\re C_{ij},\im C_{ij})$.}
\label{tab:one}
\begin{ruledtabular}
\begin{tabular}{cccc}
\backslashbox{$i$}{$j$} & $d$ & $s$ & $b$ \\
\hline
$u$ & $(0,0)$ & $(-1.6\times10^{-12},-5.2\times10^{-13})$ &
$(-3.2\times10^{-9},4.9\times10^{-9})$ \\
$c$ & $(4.5\times10^{-13},1.2\times10^{-13})$ &
$(4.9\times10^{-13},1.5\times10^{-13})$ &
$(-6.1\times10^{-8},2.1\times10^{-12})$ \\
$t$ & $(-1.5\times10^{-9},-7.9\times10^{-8})$ &
$(-1.6\times10^{-7},3.7\times10^{-7})$ &
$(-4.0\times10^{-9},1.6\times10^{-8})$ \\
\end{tabular}
\end{ruledtabular}
\end{center}
\end{table}
In Table~\ref{tab:one}, we present our results for the residual sm corrections
$C_{ij}$.
As explained in Sec.~\ref{sec:three}, in our renormalization prescription
$C_{ud}=0$.
As shown in Table~\ref{tab:one}, for the other $W\to q_i + \overline{q}_j$
amplitudes, the real and imaginary parts of $C_{ij}$ are very small.
For example, the fractional corrections of $\re C_{ij}$ with respect to the
real parts of the corresponding Born amplitude couplings, namely
$\re C_{ij}/\re V_{ij}$, reach a maximum value of $4\times10^{-6}$ for
$t\to W + s$ and are much smaller for several other cases.
It is important to note that the $C_{ij}$ are non-singular in the limits
$m_{i^\prime}\to m_i$ or $m_{j^\prime}\to m_j$, since the
$\left(m_i^2-m_{i^\prime}^2\right)^{-1}$ and
$\left(m_j^2-m_{j^\prime}^2\right)^{-1}$ singularities are canceled by the
subtraction procedure in Eq.~(\ref{eq:shift}).
For this reason, as also explained in Sec.~\ref{sec:three}, these residual
corrections can be regarded as additional finite and gauge-independent wfr
contributions, which happen to be very small.

\boldmath
\section{\label{sec:B}%
Case $(h_L)_{ii}\ne0$}
\unboldmath

Since the diagonal elements $(h_L)_{ii}$ do not contribute to the
diagonalization condition of Eq.~(\ref{eq:dis}), in the analysis of
Sec.~\ref{sec:four}, we chose $(h_L)_{ii}=0$.
We now show that the alternative choice $(h_L)_{ii}\ne0$ has no physical
effect on the $Wq_i\overline{q}_j$ coupling though ${\cal O}(g^2)$.
As explained in Sec.~\ref{sec:four}, the biunitary transformation of
Eqs.~(\ref{eq:pl}) and (\ref{eq:pr}) leads to a $Wq_i\overline{q}_j$
interaction described through terms of ${\cal O}(g^2)$ by
Eqs.~(\ref{eq:hc}) and (\ref{eq:dv}).
Writing these expressions in component form and separating out the
contributions involving the diagonal elements of $h^U$ and $h^D$, we obtain an
expression proportional to
\begin{equation}
\overline{\hat\psi}_i^{(U)}\left[V_{ij}-i\left(h_L^U\right)_{ii}V_{ij}
+V_{ij}i\left(h_L^D\right)_{jj}\right]\gamma^\lambda a_-\hat\psi_j^{(D)},
\end{equation}
which can be written as
\begin{equation}
\overline{\hat\psi}_i^{(U)}\left[1-i\left(h_L^U\right)_{ii}\right]V_{ij}
\left[1+i\left(h_L^D\right)_{jj}\right]\gamma^\lambda a_-\hat\psi_j^{(D)}
+{\cal O}(g^4).
\end{equation}
In turn, this can be expressed as
\begin{equation}
\overline{\hat\psi}_i^{(U)}\exp\left[-i\left(h_L^U\right)_{ii}\right]V_{ij}
\exp\left[i\left(h_L^D\right)_{jj}\right]\gamma^\lambda a_-\hat\psi_j^{(D)}
+{\cal O}(g^4).
\end{equation}
Since $h_L^U$ and $h_L^D$ are hermitian, the diagonal elements are real.
Thus $\exp\left[-i\left(h_L^U\right)_{ii}\right]$ and
$\exp\left[i\left(h_L^D\right)_{jj}\right]$ are multiplicative phases that can
be absorbed in redefinitions of the $\hat\psi_j^{(U)}$ and $\hat\psi_j^{(D)}$
fields.

\end{appendix}


\begin{thebibliography}{99}

\bibitem{cab} N. Cabibbo,
Phys.\ Rev.\ Lett.\ {\bf10}, 531 (1963);
M. Kobayashi and T. Maskawa,
Prog.\ Theor.\ Phys.\ {\bf49}, 652 (1973).

\bibitem{pdg}
W.-M. Yao {\it et al.}\ (Particle Data Group),
J. Phys.\ G {\bf33}, 1 (2006), and references cited therein.

\bibitem{Czarnecki:2004cw}
A. Czarnecki, W.J. Marciano, and A. Sirlin,
Phys.\ Rev.\ D {\bf70}, 093006 (2004), and references cited therein.

\bibitem{Marciano:2005ec}
W.J. Marciano and A. Sirlin,
Phys.\ Rev.\ Lett.\  {\bf96}, 032002 (2006).

\bibitem{Marciano:1975cn}
W.J. Marciano and A. Sirlin,
Nucl.\ Phys.\ {\bf B93}, 303 (1975).

\bibitem{Denner:1990yz}
A.~Denner and T.~Sack,
Nucl.\ Phys.\ {\bf B347}, 203 (1990);
B.A. Kniehl and A. Pilaftsis,
{\it ibid.}\ {\bf B474}, 286 (1996);
P. Gambino, P.A. Grassi, and F. Madricardo,
Phys.\ Lett.\ B {\bf 454}, 98 (1999);
A. Barroso, L. Br\"ucher, and R. Santos,
Phys.\ Rev.\ D {\bf62}, 096003 (2000);
Y. Yamada,
{\it ibid.}\ {\bf 64}, 036008 (2001);
K.-P.O. Diener and B.A. Kniehl,
Nucl.\ Phys.\ {\bf B617}, 291 (2001);
A. Pilaftsis,
Phys.\ Rev.\ D {\bf65}, 115013 (2002);
D. Espriu, J. Manzano, and P. Talavera,
{\it ibid.}\ {\bf66}, 076002 (2002);
Y. Zhou,
Phys.\ Lett.\ B {\bf577}, 67 (2003);
J.\ Phys.\ G {\bf30}, 491 (2004);
Y. Liao,
Phys.\ Rev.\ D {\bf69}, 016001 (2004);
A. Denner, E. Kraus, and M. Roth,
{\it ibid.}\ {\bf70}, 033002 (2004).

\bibitem{short}
B.A. Kniehl and A. Sirlin,
Phys.\ Rev.\ Lett.\ {\bf97}, 221801 (2006).

\bibitem{Feynman:1949zx}
R.P. Feynman,
Phys.\ Rev.\ {\bf76}, 769 (1949) (see especially Section~6);
Quantum Electrodynamics: A Lecture Note and Reprint Volume,
(W.A. Benjamin, Inc., New York, 1962), p.~145 {\it et seqq.}.

\bibitem{Kniehl:2000rb}
B.A. Kniehl, F. Madricardo, and M. Steinhauser,
Phys.\ Rev.\ D {\bf62}, 073010 (2000).

\bibitem{schweber}
S.S. Schweber,
An Introduction to Relativistic Quantum Field Theory,
(Row, Peterson and Company, Evanston, 1961), p.~539,
and references cited therein;
L.S.~Brown,
Phys.\ Rev.\ {\bf187}, 2260 (1969);
A. Sirlin,
Rev.\ Mod.\ Phys.\ {\bf50}, 573 (1978);
{\bf 50}, 905(E) (1978) (see Appendix~A).

\end{thebibliography}
\end{document}